\documentclass[%
 aip,
cp,  
 amsmath,amssymb,
reprint,%
]{revtex4-2}

\usepackage{graphicx}
\usepackage{dcolumn}
\usepackage{bm}

\usepackage[utf8]{inputenc}
\usepackage[T1]{fontenc}
\usepackage{mathptmx} 
\usepackage{wrapfig}

\begin{document}

\title{Baryon--anti-Baryon Photoproduction}

\author{Hao Li} 
 \email[Corresponding author: ]{hl2@andrew.cmu.edu}
\author{Reinhard A. Schumacher}%
 \email{schumacher@cmu.edu}
\affiliation{
  For the GlueX Collaboration
}
\affiliation{%
Department of Physics, Carnegie Mellon University, Pittsburgh, PA 15213, USA.
}%

\date{\today} 

\begin{abstract}
Baryon--anti-baryon photoproduction off a proton target is being studied in detail at Jefferson Lab with the GlueX Experiment. 
We observe $p\bar{p}$ and, for the first time, $\Lambda\bar{\Lambda}$ photoproduction (with $\Lambda \rightarrow \pi^- p$, $\bar{\Lambda} \rightarrow \pi^+ \bar{p}$) from thresholds up to $E_{\gamma} = 11.4$ GeV. 
Preliminary spectra from data accumulated during the GlueX Phase-I period are shown. 
Angular distributions of the photoproduced hyperons indicate more than one production mechanism in the reaction channel $\gamma p \rightarrow \Lambda\bar{\Lambda}p$.
A Monte Carlo simulation with four mechanisms, tested through comparison between simulation and experimental data, is presented. 
\end{abstract}

\maketitle

\section{INTRODUCTION}
The GlueX experiment at the Thomas Jefferson National Accelerator Facility is dedicated to better understand Quantum Chromodynamics (QCD), in particular the nature of quark confinement in light-quark mesons and baryons.
Taking advantage of GlueX data, we are studying the following two reactions concerning the interactions between nucleons and hyperons:
\begin{align}
\gamma p &\rightarrow p\bar{p} p \\
\gamma p &\rightarrow \Lambda\bar{\Lambda} p 
\end{align}

We are interested in a comparison of the two reactions, 
given their presumed similar production mechanism and the absence/presence of the strange quark.
The result may shed light not only on the interaction potentials in the nucleon-nucleon system ($p\bar{p}$), 
but also on the strange-strange baryonic system ($\Lambda\bar{\Lambda}$), 
and furthermore, 
the strange--nonstrange baryonic system ($p\bar{\Lambda}$).

\begin{wrapfigure}{r}{0.4\textwidth}
	\vspace{-20pt}
	\begin{center}
		\includegraphics[width=0.38\textwidth]{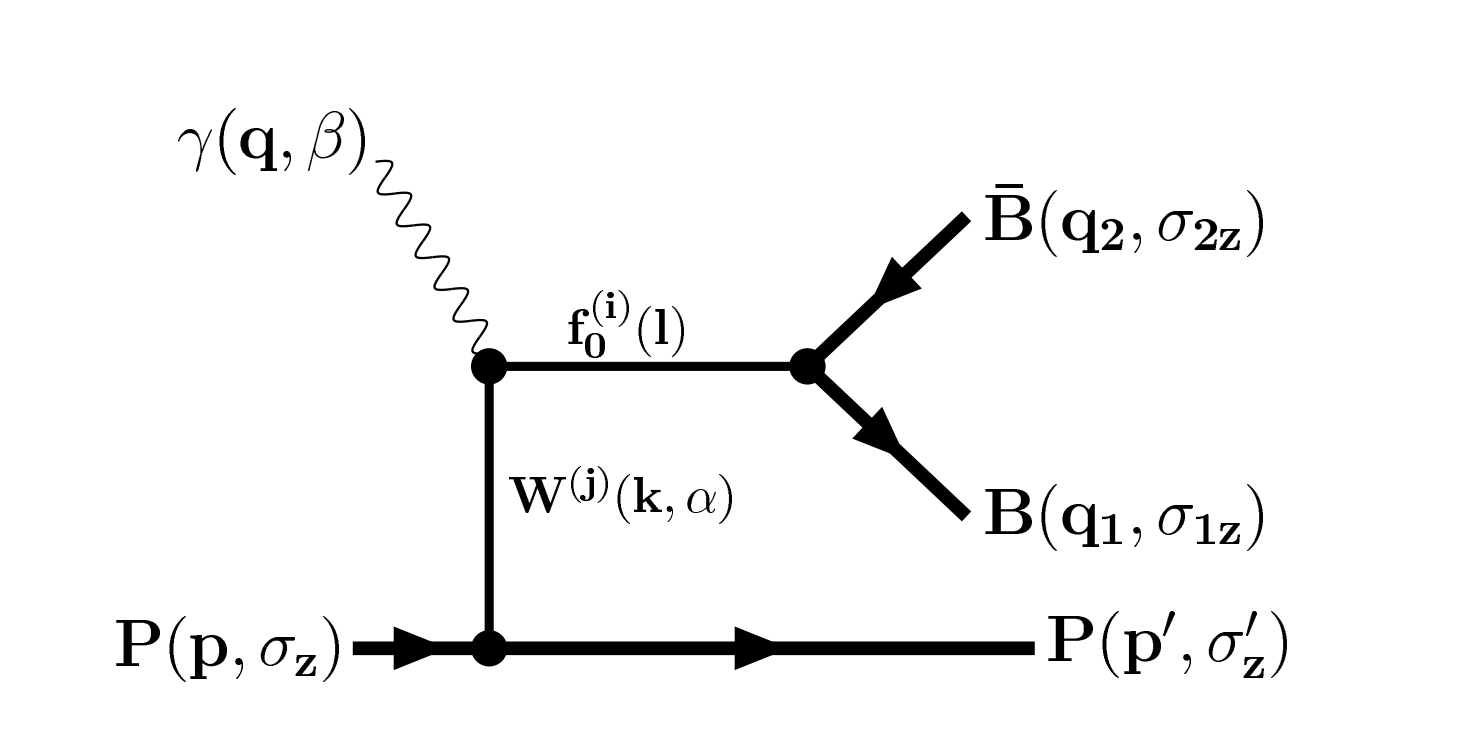}
	\end{center}
		\vspace{-10pt}
	\caption{\label{fig:theoretical} One possible route for photoproduction of baryon--anti-baryon pairs as observed in GlueX. The Feynman diagram here describes the contribution of the intermediate scalar mesons $f_0^{(i)}=f_0(1370), f_0(1500)$ and $f_0(1710)$ to the photoproduction of the $B\bar{B}$ pair, through t-channel exchange of both vector meson $V^{(j)}=\rho^0,\omega$ and axial-vector meson $A^{(j)}=b_1,h_1$ as well as Reggeons. Note that here $W=V,A$ and $B=p,\Lambda$.   The actual reaction mechanism is presently unknown.}
\end{wrapfigure}

Little relevant theoretical work has been done on $p\bar{p}$ and $\Lambda\bar{\Lambda}$ photoproduction. 
In the energy range of the GlueX experiment, one expects to see the dominant t-channel exchange of some mesons, Pomerons or Reggeons leading to the $p\bar{p}$ final state. 
One recent theoretical effort \cite{Gutsche:2017xtm} to model the baryon--anti-baryon pair photoproduction is illustrated in Fig.~\ref{fig:theoretical}.
It assumes that the pair is the decay product of some intermediate scalar mesons at the upper vertex. 
According to this model, both vector mesons and axial-vector mesons as well as the Reggeons are allowed in the t-channel exchange. 
Aside from the different couplings to the final state baryons, this model predicts very similar t-dependence of the cross section and the beam spin asymmetry, $\Sigma$, for both reactions. The ongoing experimental measurement will be compared with those predictions to check the validity of the proposed theoretical model. 
In addition, as all proposed intermediate scalar mesons are below threshold of the two reactions, it may be possible to identify intermediate states of higher masses, as well as production mechanisms different than t-channal exchange.

Recently, the BES-III collaboration reported observations of near-threshold enhancement in the radiative decay of $J/\Psi\rightarrow\gamma\{p\bar{p}\}$ \cite{bai2003observation} and the decay of $J/\Psi\rightarrow\{\Lambda\bar{\Lambda}\}$ \cite{ablikim2018observation}. Because the two reactions we investigate feature similar final states, we seek any unexpected phenomenon observed in the corresponding near-threshold regions. 
There has been a long history at CERN and other facilities, searching for ``baryonium'' in the $p\bar{p}$ system \cite{Klempt:2002ap}, without any evidence of continuum quasi-bound states found. 
This makes the BES-III result more interesting and adds motivation to examining the related channels at GlueX.

Reaction (1) was measured by the CLAS experiment at Jefferson Lab with beam energy up to about 5.5 GeV \cite{Phelps}. 
With GlueX, we will be able to expand the beam energy up to 11.4 GeV with higher statistics.
Very preliminary results for Reaction (2) were presented previously \cite{Schumacher:2018xnh}.

\section{APPARATUS AND PROCEDURES}

\begin{figure}[ht]
		  \centering
		  \begin{minipage}[b]{0.42\textwidth}
		    \includegraphics[width=\textwidth]{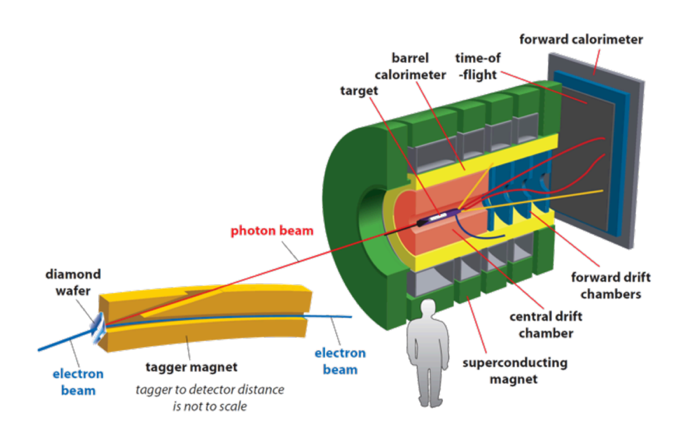}
		    \caption{\label{fig:gluex_detector} Schematic of the GlueX detector and the photon beam tagger.}
		  \end{minipage}
		  \hspace{1.5cm}
		  \begin{minipage}[b]{0.42\textwidth}
		    \includegraphics[width=\textwidth]{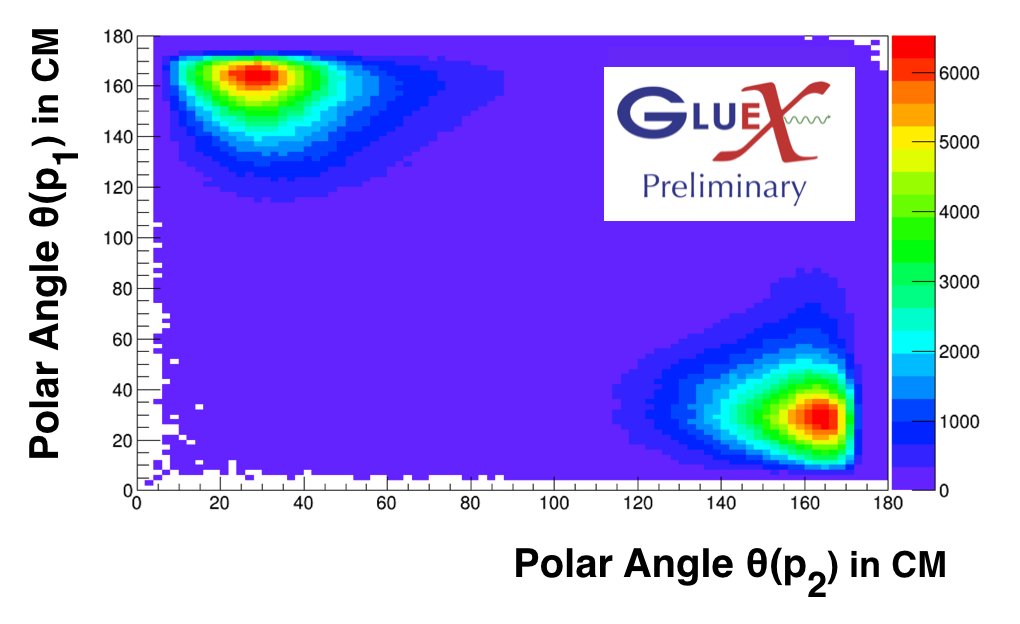}
		    \caption{\label{fig:ppbar_angle} The angular correlation between the two detected protons in the center-of-mass frame for $\gamma p \rightarrow p_1 p_2 \bar{p}$. }
		  \end{minipage}
		\end{figure}

	The GlueX detector in Hall D at Jefferson Lab is illustrated in Fig.~\ref{fig:gluex_detector}. 
	An 11.6 GeV electron beam, delivered by the CEBAF (Continuous Electron Beam Accelerator Facility), is incident on a diamond radiator and converted to a “coherent” Bremsstrahlung beam of photons. 
	The photon beam tagger uses a dipole magnet and an array of fine-grained hodoscope scintillators to measure the recoiling electrons in order to compute the energy of the beam photons.
	After collimation, the degree of polarization in the coherent peak (near 9 GeV) is about $40\%$. 

	The photon beam strikes a cylindrical LH2 target of 30 cm length. The target is surrounded by a Start Counter made of scintillators. 
	A Central Drift Chamber with axial geometry surrounds the Start Counter to track charged particles, together with a planar Forward Drift Chamber placed in the downstream direction. Outside the CDC and FDC, a lead-scintillator Barrel Calorimeter is responsible for detecting mainly neutral particles.
	These parts of the GlueX detector are placed in a 2 T solenoidal magnetic field.
	Downstream there is a Time-of-Flight scintillator wall and an electromagnetic Forward Calorimeter. 

	As a fixed-target experiment in the multi-GeV energy range, most of the reaction products are boosted forward in the laboratory frame. The detector is almost $4\pi$ hermetic, covering charged particle tracking in polar angle from $2^{\circ}$ to $128^{\circ}$. The momentum resolution for charged particles is $\sigma(p)/p\approx 1 - 5\%$ for reconstructed pions above 100 MeV/c and protons above 300 MeV/c.

	The experiment has been accumulating data since the Sping of 2016. The GlueX Phase-I program has been completed with four data sets taken during Spring 2016, Spring 2017, Spring 2018, and Fall 2018. About 0.4 million $\Lambda\bar{\Lambda}p$ events and 12 million $p\bar{p}p$ events have been collected.
	Events are selected from reconstructed tracking and calorimeter information according to the final state topology of the corresponding reaction hypotheses. 
	A kinematic fit with momentum, vertex and mass constraints is applied, together with particle identification (PID) cuts, to identify and reject background events.

	One challenge in analyzing the $p_1 p_2 \bar{p}$ final state is that two indistinguishable protons ($p_1$ and $p_2$, arbitrarily ordered) are present. Their respective roles as target or produced particles must be inferred from the data, and this cannot be done without some ambiguity.

	As shown in Fig.~\ref{fig:ppbar_angle}, it appears that at the GlueX energies ($5.5$ GeV$<E_{\gamma}<11.6$ GeV) the polar angles of the two protons are highly correlated in the center-of-mass frame, i.e. for any single event, one proton goes very forward and one goes very backward. 
	Thus, we sort the two protons of each event based on their CM polar angles, i.e. assigning ``forward proton'' ($p_{fwd}$), and ``backward proton'' ($p_{bkwd}$).

		\begin{figure}[ht]
		  \centering
		  \begin{minipage}[b]{0.3\textwidth}
		    \includegraphics[width=\textwidth]{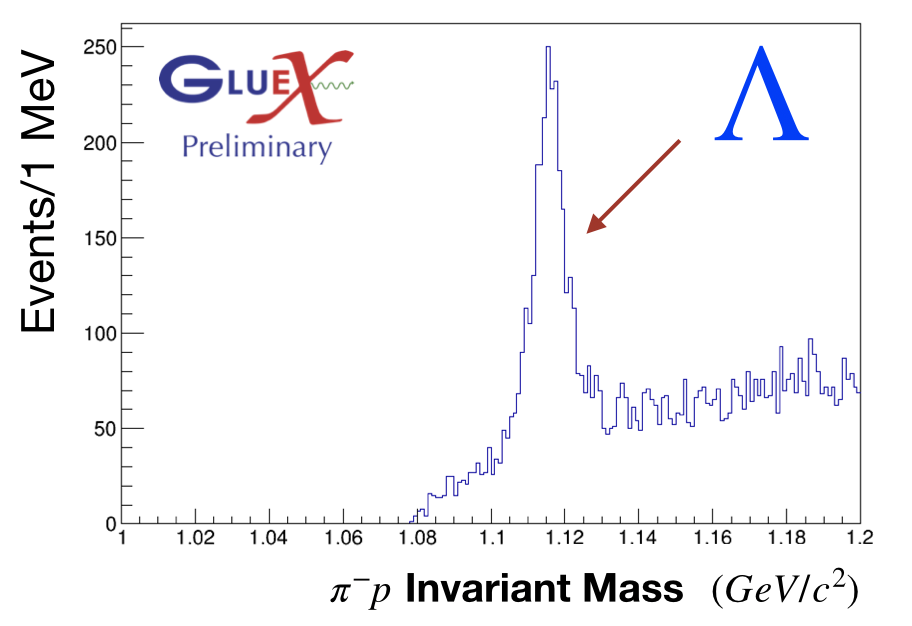}
		  \end{minipage}
		  \begin{minipage}[b]{0.3\textwidth}
		    \includegraphics[width=\textwidth]{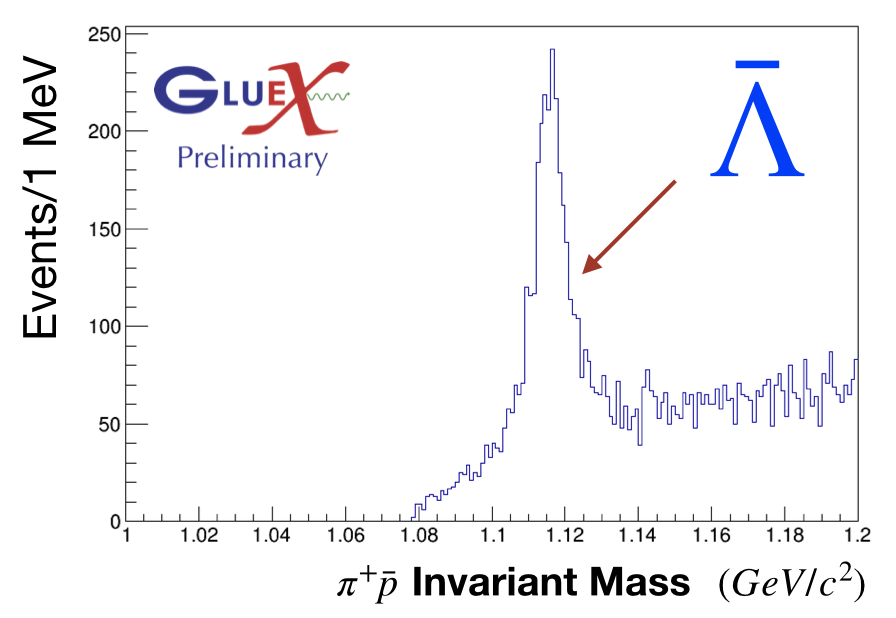}
		  \end{minipage}
		  \begin{minipage}[b]{0.38\textwidth}
		    \includegraphics[width=\textwidth]{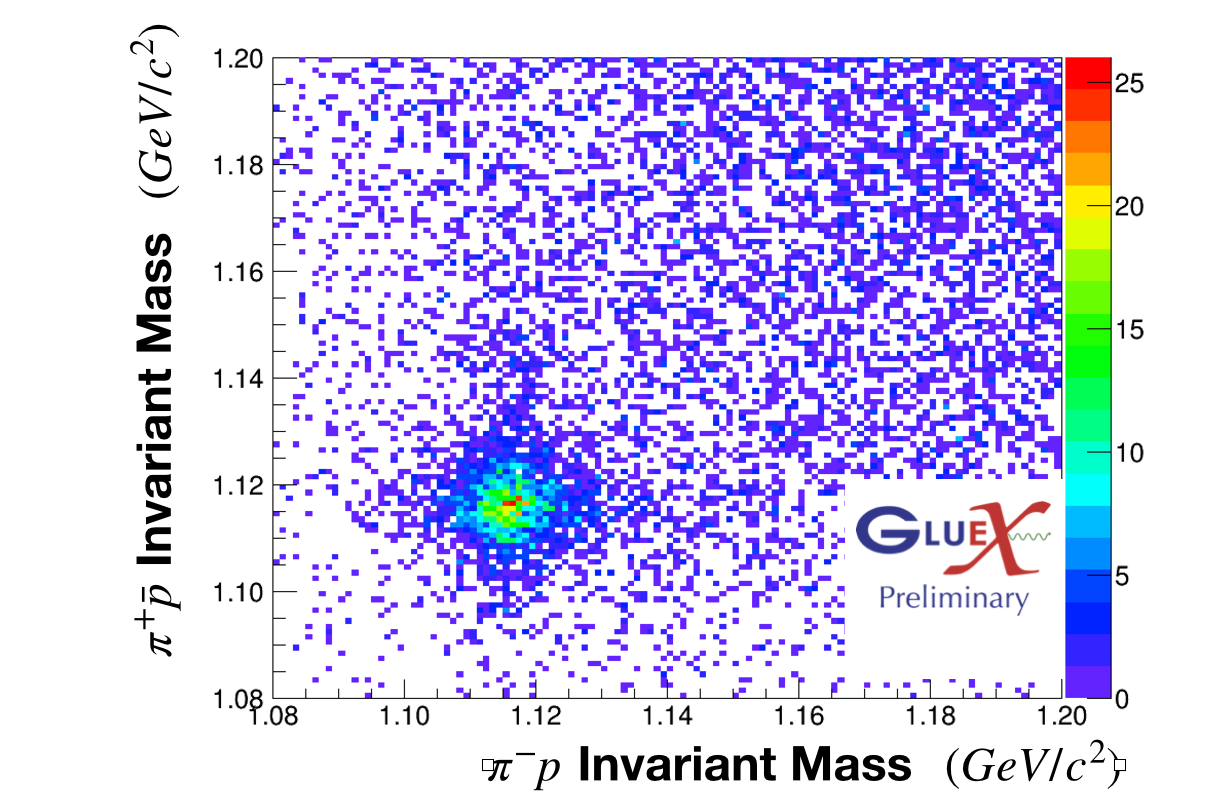}
		  \end{minipage}
		  \caption{\label{fig:lamlambar_IM} The invariant mass of the reconstructed $\Lambda$ (left) from the detected $\pi^-$ and $p$, and the corresponding distribution for the $\bar{\Lambda}$ (middle) reconstructed from $\pi^+$ and $\bar{p}$. The kinematic fit used to produce these distributions did not constrain the hyperon masses. The correlation between the invariant masses of the hyperon pairs is shown on the right.  }
		\end{figure}

	For the $\Lambda\bar{\Lambda}p$ channel, the distributions of the reconstructed $\Lambda \rightarrow \pi^- p$ and $\bar{\Lambda} \rightarrow \pi^+ \bar{p}$ invariant masses are presented in Fig.~\ref{fig:lamlambar_IM}. 
	To avoid bias, the hyperon masses were not constrained in this kinematic fit. As shown in the first two panels, the hyperon pairs produce clear mass signals in the GlueX detector. 
	The right panel shows the correlation between $\Lambda$ and $\bar{\Lambda}$. 
	There is a smooth background that can be removed by improved pion rejection in the proton signal at high momenta and by a sideband subtraction technique.
	For the present work, $\Lambda\bar{\Lambda}$ events can be further selected using a circular cut in the correlated mass distribution.

\section{PRELIMINARY RESULTS}

	\subsection{Angular Distributions}
	Some preliminary angular distributions in the center-of-mass frame are presented in this section for the two reactions. The lineshapes of those angular distributions reveal evidence of more than one production mechanism.

		\begin{figure}[ht]
			\begin{center}
				\includegraphics[width=0.65\textwidth]{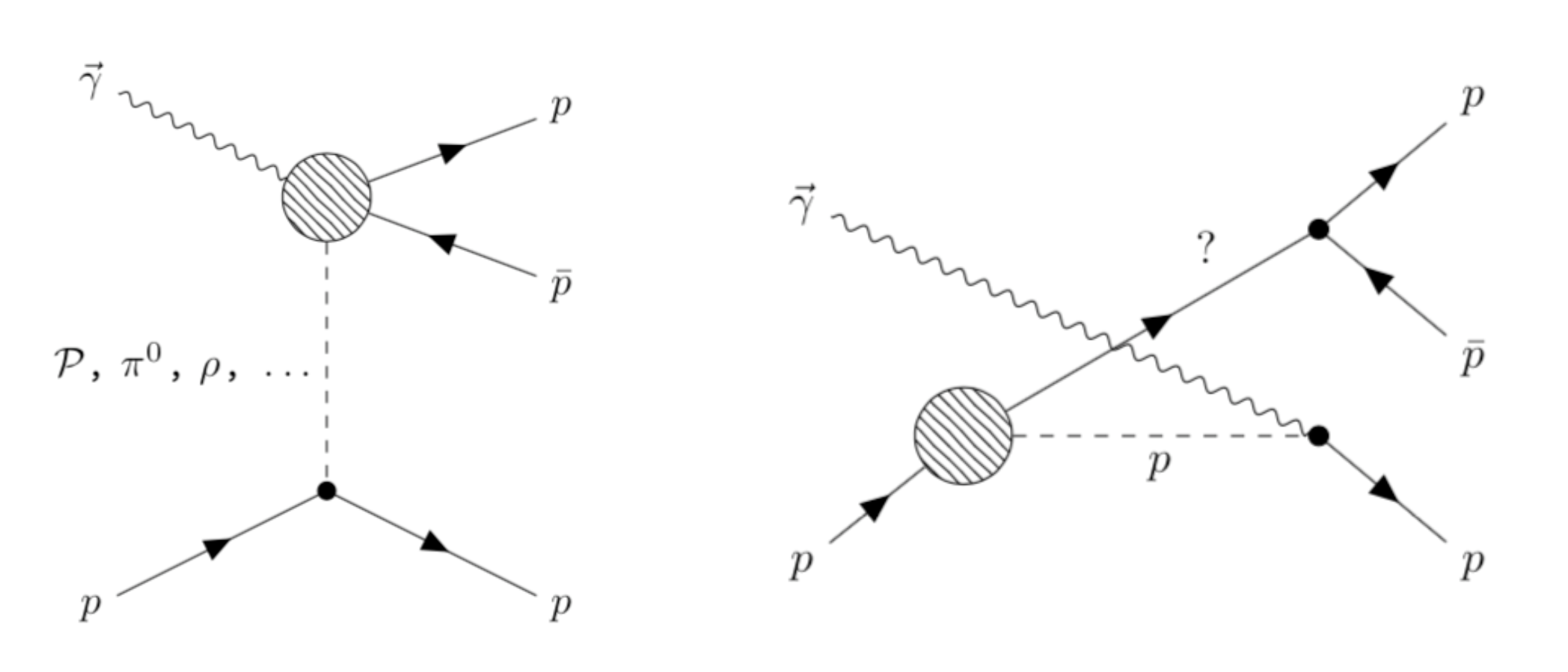}
				\caption{\label{fig:ppbar_reaction} The Feynman diagram shown the t-channel (left) and u-channel (right) production mechanisms of the reaction $\gamma p \rightarrow p \bar{p} p$. }
			\end{center}
		\end{figure}
\begin{figure}[ht]
		\centering
		  \begin{minipage}[b]{0.3\textwidth}
		    \includegraphics[width=\textwidth]{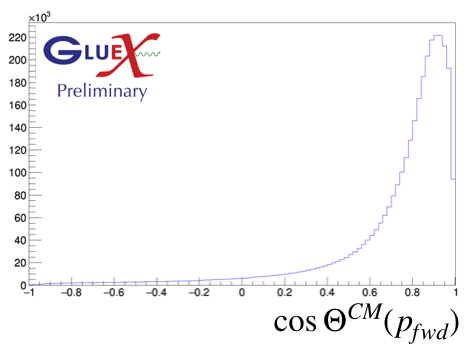}
		  \end{minipage}
		  \begin{minipage}[b]{0.315\textwidth}
		    \includegraphics[width=\textwidth]{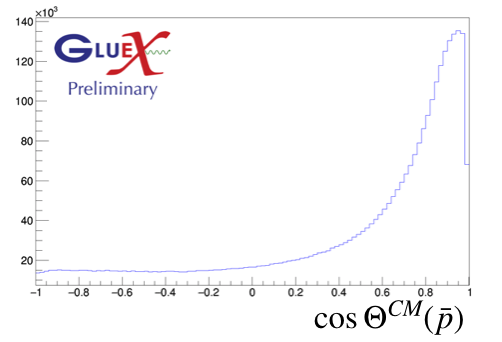}
		  \end{minipage}
		  \begin{minipage}[b]{0.3\textwidth}
		    \includegraphics[width=\textwidth]{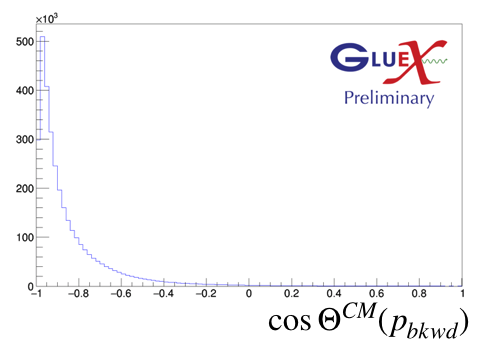}
		  \end{minipage}
		\caption{\label{fig:ppbar_anglular} The angluar distributions of the two protons (left, right), and the antiproton (middle) in the $\gamma p \rightarrow p \bar{p} p $ reaction channel in the overall center-of-mass frame. }
		\end{figure}
\begin{figure}[ht]
		\centering
		  \begin{minipage}[b]{0.3\textwidth}
		    \includegraphics[width=\textwidth]{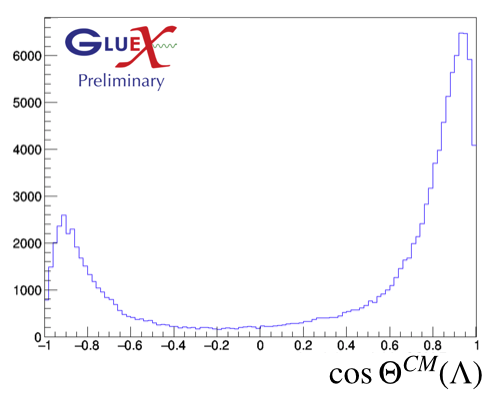}
		  \end{minipage}
		  \begin{minipage}[b]{0.307\textwidth}
		    \includegraphics[width=\textwidth]{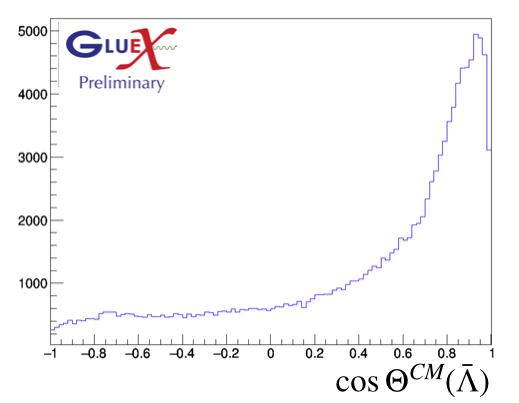}
		  \end{minipage}
		  \begin{minipage}[b]{0.3\textwidth}
		    \includegraphics[width=\textwidth]{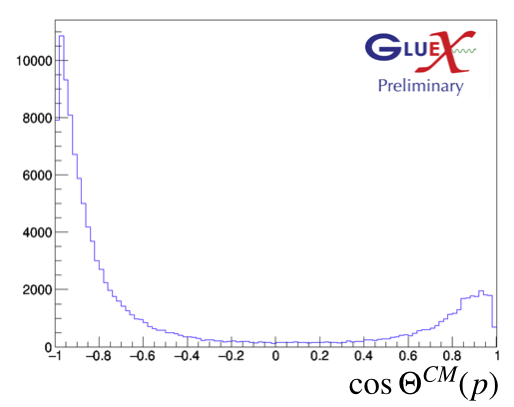}
		  \end{minipage}
		\caption{\label{fig:lamlambar_anglular} The angular distributions of the three particles in the $\gamma p \rightarrow \Lambda \bar{\Lambda} p $ reaction channel in the overall center-of-mass frame. }
		\end{figure}

	It is expected to have multiple pathways towards baryon--anti-baryon photoproduction's final state. 
	As shown in Fig.~\ref{fig:ppbar_reaction}, the reaction $\gamma p \rightarrow p\bar{p}p$ can produce $p\bar{p}$ via both t-channel and u-channel exchange mechanisms, 
	while s-channel production is suppressed at GlueX energies. 
	An unknown intermediate state is conjectured that decays into the $p\bar{p}$ pair.
	Assuming that the pair decays isotropically from an intermediate state in its own pair-rest frame (due to parity conservation), the angular distributions of the two particles should be exactly the same. 
	In the overall center-of-mass frame, it is expected to see t-channel produced pairs going in the forward direction, while the other particle goes in the opposite direction (due to total momentum conservation). 
	The GlueX data confirms that there are symmetric forward-going peaks in both baryon's and anti-baryon's angular distributions, as shown on the left and middle panels in Fig.~\ref{fig:ppbar_anglular} and Fig.~\ref{fig:lamlambar_anglular}. Furthermore, the peak in the opposite direction is observed, as shown on the right panel. 
	
	However, something else is observed in addition to the symmetric event distributions.  
	As shown on the middle panel of the Fig.~\ref{fig:ppbar_anglular}, in the $p\bar{p}p$ channel, $\bar{p}$ has an extra broader distribution in addition to the dominant forward peak. 
	This signals some additional u-channel contribution.
	In the $\Lambda\bar{\Lambda}p$ channel, a similar broad distribution of the anti-baryon is also observed (middle panel of the Fig.~\ref{fig:lamlambar_anglular}).
	In addition, in the $\Lambda\bar{\Lambda}p$ channel a backward proton peak and a corresponding $\Lambda$ forward peak is observed (left and right panels), respectively.
	This implies that several other reaction mechanisms may be present, other than the dominant t-channel pair-production picture. 
	The angular spectra shown in this article have not been background subtracted, nor corrected according to the detector acceptance.

	\subsection{Reaction Mechanisms}

	In a three-body final state, there are usually some simple reaction models to help understand the process of the reaction.
	\begin{wrapfigure}{l}{0.30\textwidth}
			\vspace{-25pt}
			\begin{center}
				\includegraphics[width=0.28\textwidth]{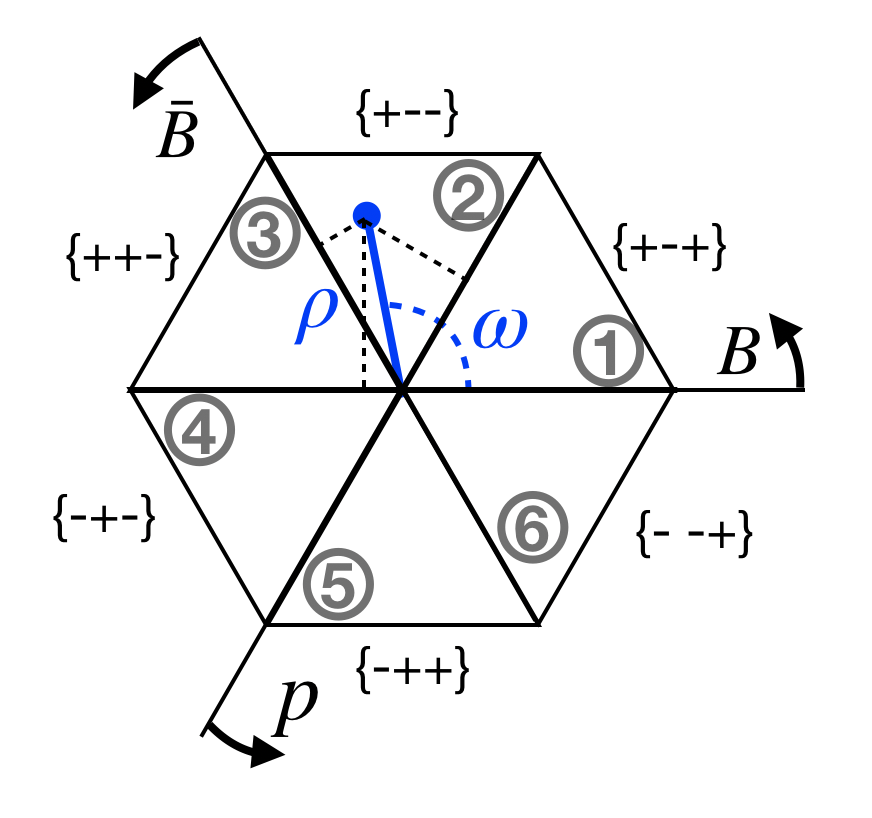}
				\caption{\label{fig:vanhove}
		  		The van Hove plane with one data point (blue). The distances from the point to the three axes are the three longitudinal momenta. The six sectors are indexed counter-clockwisely. }
			\end{center}
			\vspace{-20pt}
	\end{wrapfigure}
	 
	A pure phase space model assumes no interaction between the particles. 
	While constrained by the conservation of total momentum and energy, the three particles decay isotropically from the main reaction vertex. 
	This model can be used to learn the kinematic boundaries and phase space factors that bound the more realistic models that we describe in this section.

	In our context, we focus on the process that either the beam photon or the target proton (also referred as t- or u- channel in this article) becomes an off-shell intermediate state after some particle exchange.
	Depending on the exchange mechanism, 
	the intermediate state can decay into various combinations of two final state particles, while the third particle in the final state goes into the opposite direction in the center-of-mass frame to conserve momentum. 
	This conjecture leads to a combinatoric problem of the three particles in the final state. The van Hove diagram \cite{van1969final,van1969longitudinal,bialas1969longitudinal} is effective to examine this problem by investigating the longitudinal momentum space of the final states.

	A van Hove diagram is a system that can be displayed on a graph with three axes spaced $\frac{2\pi}{3}$ apart in an adaptation of the well-known Viviani theorem of geometry.
	For a three-body system including a baryon ($B$), an anti-baryon ($\bar{B}$), and a proton (p), let $p_z^{i}$ be the center-of-mass longitudinal momentum of particle $i$. 
	As shown in Fig.~\ref{fig:vanhove}, the longitudinal momenta of any three-body final state occupies a unique point in this graph. 
	The conservation of longitudinal momentum in the center-of-mass frame gives $p_z^{B} + p_z^{\bar{B}} + p_z^{p} = 0$, which means two longitudinal momenta should have signs opposite to the third.
	The plane is divided into six sectors, corresponding to six possible combinations of the signs (+ - +, + - -, + + -, - + -, - + +, - - +, as shown in Fig.~\ref{fig:vanhove} as Sectors 1-6).
	The longitudinal momenta are parameterized by a radius $\rho$ and a polar angle $\omega$ (explicit representation written in the appendix) with sinusoidal functions to map the signs into phase space:

	\begin{align}
		p_z^{B}        &= \rho \sin(\omega)\\
		p_z^{\bar{B}}  &= \rho \sin(\omega-\frac{2}{3}\pi)\\
		p_z^{p}        &= \rho \sin(\omega-\frac{4}{3}\pi)
	\end{align}

		\begin{figure}[hbt]
		\includegraphics[scale=0.42]{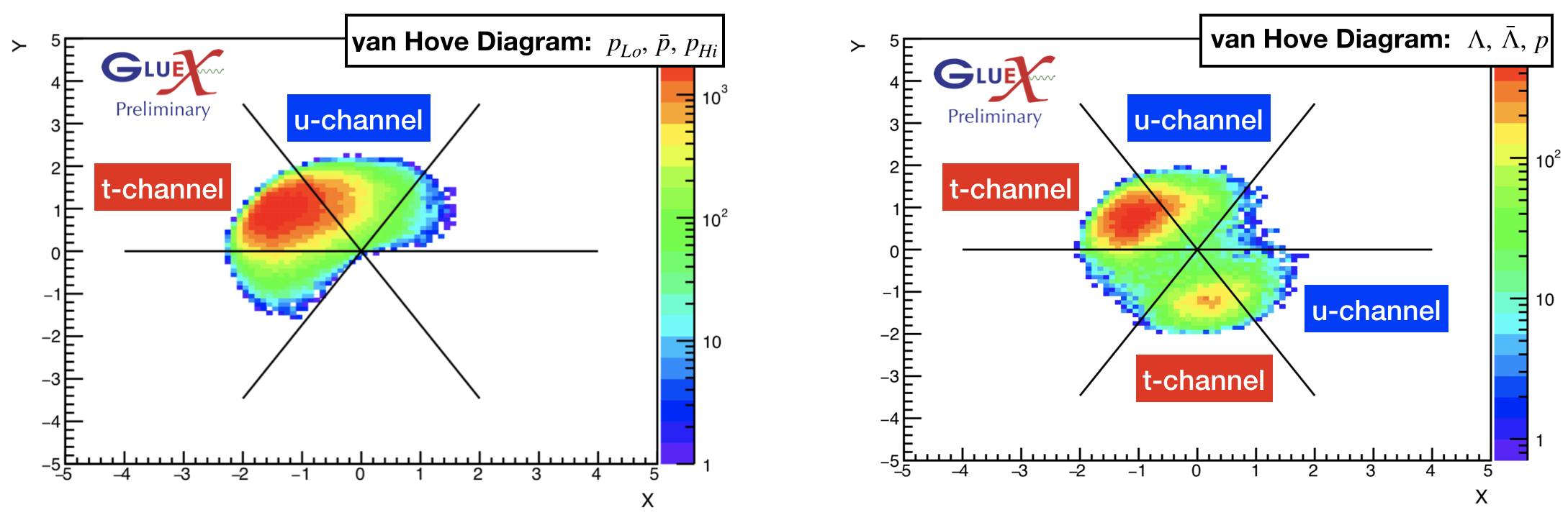}
		\caption{\label{fig:vhplots} 
		(Left) The van Hove diagram for the $\gamma p \rightarrow p_{fwd} ~ \bar{p} ~ p_{bkwd}$ reaction channel. (Right) The van Hove diagram for the $\gamma p \rightarrow \Lambda \bar{\Lambda} p$ reaction channel. Different exchange mechanisms are recognized and tagged on each individual sectors. Data have been summed over the GlueX beam energy range 5.5 GeV $< E_{\gamma}<$ 11.6 GeV.}
		\end{figure}

	  	\begin{wrapfigure}{r}{0.58\textwidth}
		\vspace{-20pt}
		\begin{center}
			\includegraphics[width=0.54\textwidth]{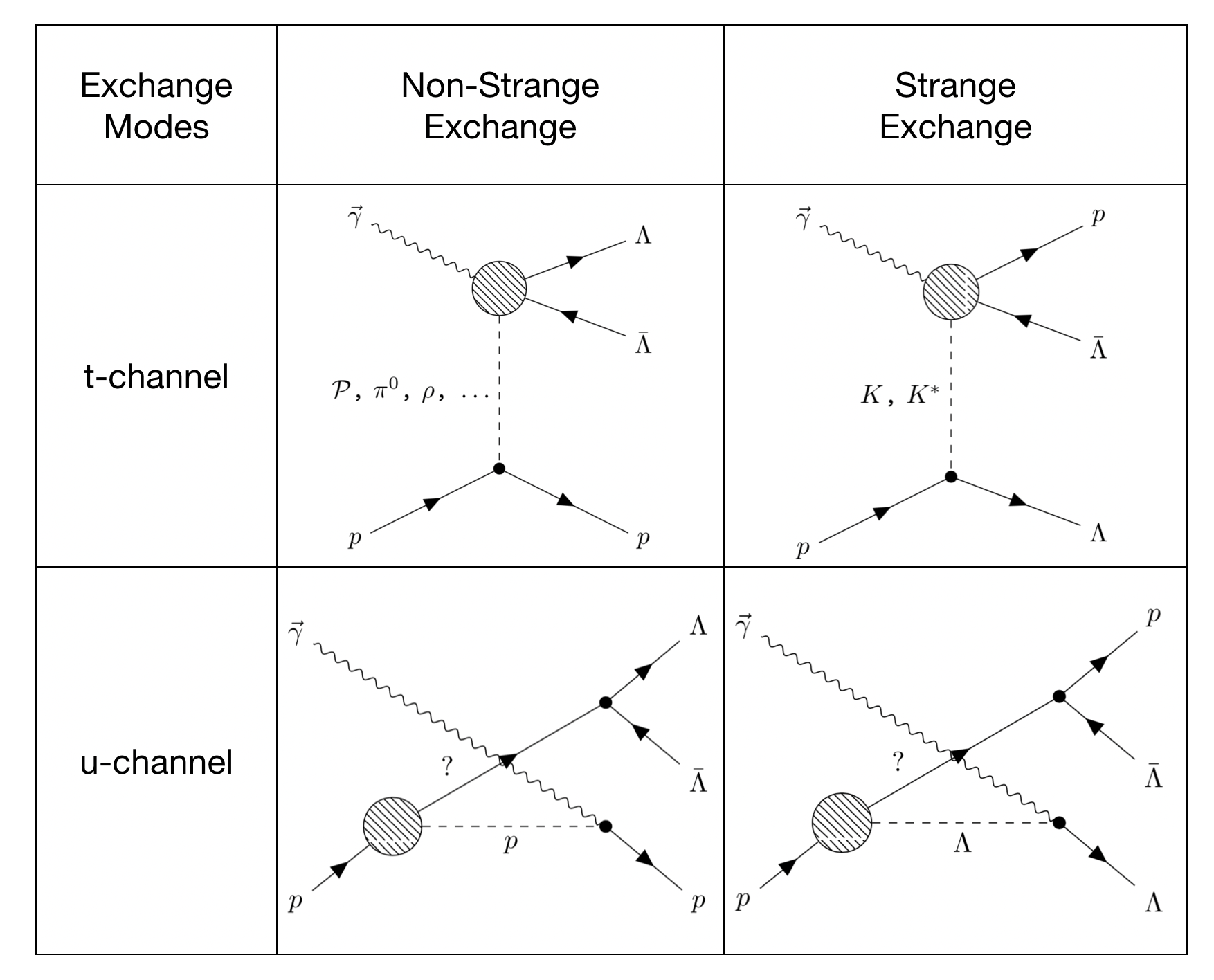}
			\caption{\label{fig:table} 
			Different production mechanisms categorized by the exchange particles for the reaction $\gamma p \rightarrow \Lambda\bar{\Lambda}p$.}
		\end{center}
		\vspace{-20pt}
	 \end{wrapfigure}
	 As shown on the left in Fig.~\ref{fig:ppbar_reaction}, we assume that the reaction $\gamma p \rightarrow \{p\bar{p}\} p$ can occur via two possible exchange mechanisms. 
	 In the t-channel scenario, the $p\bar{p}$ produced at the upper vertex carries most of the momentum from the incoming beam photon, while the proton at the lower vertex acquires little longitudinal momentum in the exchange. 
	 The signs of longitudinal momenta for $p_{fwd} ~ \bar{p} ~ p_{bkwd}$  are then $(++-)$ , when boosted to the center-of-mass frame.  
	 Analogously, in the u-channel scenario, the proton produced at the right vertex carries most of the momentum from the incoming beam photon, while the intermediate state produced at the left vertex acquires little from the photon momentum hence the $p\bar{p}$ pair is produced more backward. The signs of longitudinal momenta are then $(+--)$.
	 This is shown for GlueX data in Fig.~\ref{fig:vhplots} (left), where the data distribution occupies Sectors 1-4. Sectors 2 and 3 are recognized as u-channel and t-channel event distributions. For Sectors 1 and 4, where the $pp$ system pair production is dominant, no evidence of such events is observed, other than the dominant distributions leaking from the neighboring sectors.

	 In the reaction $\gamma p \rightarrow \Lambda \bar{\Lambda} p$, both non-strange and strange particles can be exchanged. 
	 This is evident in the right hand panel of Fig.~\ref{fig:vhplots}, in contrast to the left hand panel for the $\gamma p \rightarrow \{p\bar{p}\} p$ reaction. 
     According to GlueX data, evidence of a co-moving $p\Lambda$ system is not found in Sectors 1 and 4. Sectors 3 and 6 are recognized as t-channel and u-channel $\Lambda \bar{\Lambda}$ pair production via exchange of non-strange particles; Sectors 2 and 5 are recognized as t-channel and u-channel $p \bar{\Lambda}$ pair production via exchange of strange particles. 

	 The four hypothetical exchange mechanisms are summarized in a table (Fig.~\ref{fig:table}). The mechanisms in the first column produce $\Lambda\bar{\Lambda}$ pairs through exchange of non-strange mesons or Pomerons. The mechanisms in the second column produce $p \bar{\Lambda}$ pairs through exchange of kaons or hyperons.

	\subsection{Monte Carlo Simulations}

	Given the production hypotheses (Fig.~\ref{fig:ppbar_reaction}, Fig.~\ref{fig:table}) as identifiable from the event distributions in van Hove diagrams (Fig.~\ref{fig:vhplots}), those reaction mechanisms have been simulated using the GlueX GEANT4-based simulation program. The simulations can be tuned to approximate GlueX data with a good level of agreement. 
	The comparison between different exchange mechanisms may help gain deeper understanding into the nature of the interaction potentials between not only nucleon-nucleon system, but also nucleon-hyperon system and hyperon-hyperon system.

		\begin{wrapfigure}{r}{0.42\textwidth}
			\vspace{-20pt}
			\begin{center}
				\includegraphics[width=0.4\textwidth]{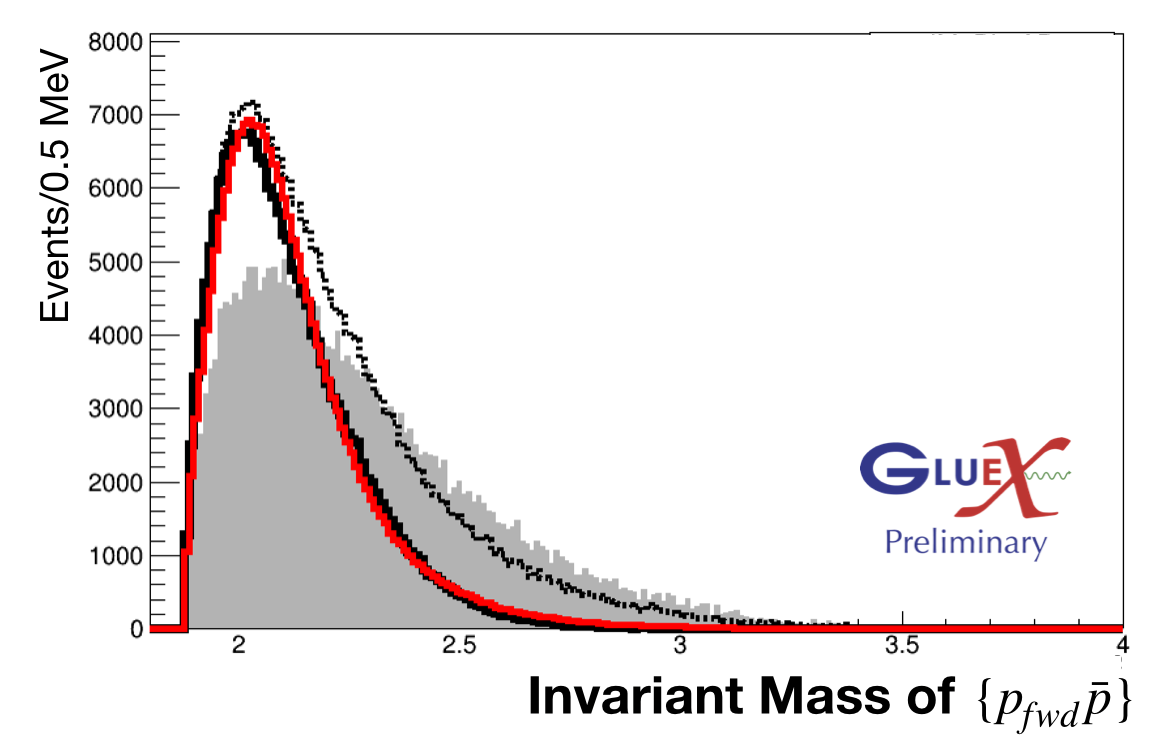}
				\caption{\label{fig:pp_t} Invariant mass distributions of $p_{fwd}\bar{p}$ are shown here. The black dashed line is the GlueX data of all mechanisms. The black solid line is the GlueX data with the van Hove angle between $120^{\circ}$ and $180^{\circ}$, where t-channel is expected to be the dominant mechanism. The red solid line is the simulation in t-channel with the van Hove angle between $120^{\circ}$ and $180^{\circ}$. The grey filled distribution is the 3-body phase space model in the same van Hove angle range. }
			\end{center}
			\vspace{-10pt}
		\end{wrapfigure}

	The simulated reactions are generated using a customized event generator.
	The beam energy distribution in the event generater is extracted from the GlueX experiment.
	The reaction is simulated as either t- or u- channel, according to the selected momentum transfer, with a single Regge parameter $b_{R}$ to characterize the exchange strength. 
	At one production vertex, an intermediate state with chosen mean mass $\texttt{m}_{\texttt{BW}}$ and resonance width $\Gamma_{\texttt{BW}}$ is produced, subsequently decaying into a baryon--anti-baryon pair. 
	In the rest frame of this pair, the two particles are arranged to decay isotropically. 
	The simulated events were processed in the same way as real physics data. 

	The comparison between GlueX data and the t-channel Monte Carlo simulation of the $\{p\bar{p}\}p$ reaction is shown in Fig.~\ref{fig:pp_t}. 
	To eliminate the distraction from the u-channel background in GlueX data, a selection on the van Hove angle is applied to select on the sectors where the t-channel events are dominant. 
	As shown in the comparison, there is a good agreement between data and the t-channel simulation in the van Hove angle range from $120^{\circ}$ to $180^{\circ}$, where the antiproton and one of the protons go forward in the CM frame.
	In addition, comparison can be made to a pure three-body phase space picture with the same set of cuts as the real events.
	The comparison shows that the real data invariant mass spectra rise up much faster from the $p\bar{p}$ threshold, suggesting some attractive potential between the proton and the anti-proton. 
	Here we refrain from specification of the $p\bar{p}$ interaction model since these results are very preliminary.
	
		\begin{figure}[ht]
		\includegraphics[scale=0.4]{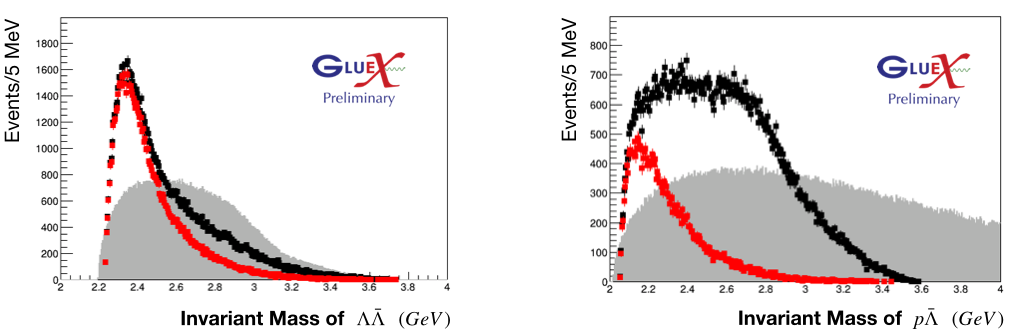}
		\caption{\label{fig:IM_lamp}  
		Left panel is the invariant mass distributions of the $\Lambda\bar{\Lambda}$ system. GlueX events selected in the van Hove angle range $0^{\circ}<\omega < 210^{\circ}$ (red points) are compared to the full dataset (black dots);
		Right panel is the invariant mass distributions of the $p\bar{\Lambda}$ system. Events selected from $210^{\circ}<\omega<360^{\circ}$ (red dots) are compared to the full dataset (black points). 
		In both panels, invariant mass distributions of the corresponding systems in three-body phase space are shown in grey filled regions. }
		\end{figure}

	There are two dominant production mechanisms of the final state $\Lambda \bar{\Lambda} p$. 
	In Fig.~\ref{fig:vhplots} (right), those two mechanisms appear as event distributions dominating Sector 3 (t-channel $\Lambda \bar{\Lambda}$ pair production) and Sector 5 (t-channel $p \bar{\Lambda}$ pair production). 
	A straightfoward way to separate those two mechanisms is with a van Hove angle cut at $\omega = 210^{\circ}$. 
	However, the neighboring u-channel backgrounds are difficult to eliminate, which may eventually require further multivariate analysis. 
	In Fig.~\ref{fig:IM_lamp}, the invariant masses of the $\Lambda \bar{\Lambda}$ system (left) and the $p \bar{\Lambda}$ system (right) are shown. The events are extracted from the GlueX data for van Hove angle $\omega < 210^{\circ}$ and $\omega > 210^{\circ}$, respectively. 
	Both invariant mass distributions are compared to the full dataset ($0^{\circ}<\omega< 360^{\circ}$) to show the effect of the van Hove angle selection. 
	On the left panel, we believe that the events with van Hove angle above $210^{\circ}$ involve no interaction between the two hyperons, therefore their contribution to the full dataset distribution is different from the signal events selected.
	On the right panel, the events with van Hove angle lower than $210^{\circ}$ have no interaction between $\bar{\Lambda}$ and $p$, therefore their contribution to the full dataset distribution is different from the signal events selected.

	The selected invariant mass distributions are also compared to the pure three-body phase space model. The difference in how the two lineshapes rise up right after the $\Lambda \bar{\Lambda}$ and $p \bar{\Lambda}$ thresholds may imply the attractive potentials in the system.

\begin{figure}[ht]
		  \centering
		  \begin{minipage}[b]{0.42\textwidth}
		    \includegraphics[width=\textwidth]{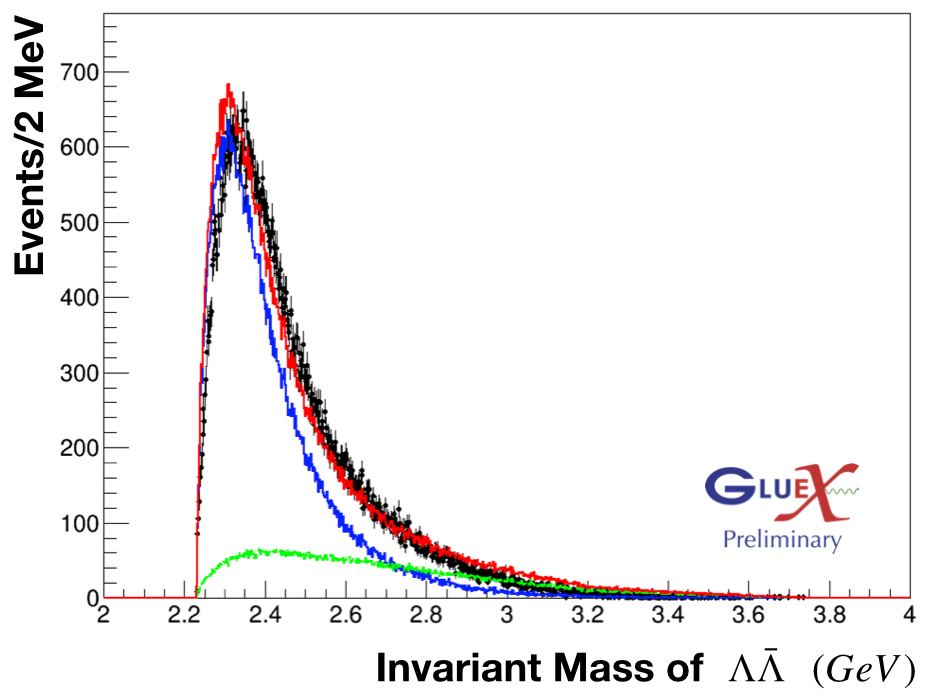}
		  \end{minipage}
		  \hspace{1cm}
		  \begin{minipage}[b]{0.43\textwidth}
		    \includegraphics[width=\textwidth]{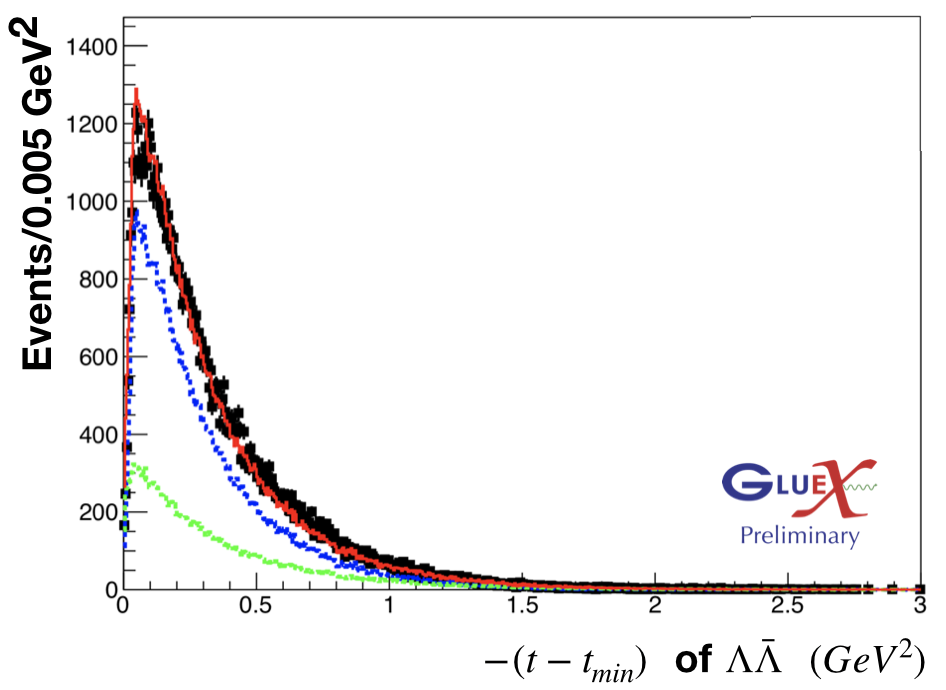}
		  \end{minipage}
		  \caption{\label{fig:lamlambar_MC} 
		  (Left)  The invariant mass of the $\Lambda\bar{\Lambda}$ system. 
		  (Right) The $t-t_{min}$ distribution. 
		   GlueX events selected in the van Hove angle range $0^{\circ}<\omega < 210^{\circ}$ (black points) are compared to the combined MC (red lines). The combined MC is the sum of t-channel MC (blue lines) of $\Lambda \bar{\Lambda}$ pair production and the u-channel MC (green lines) of the $p \bar{\Lambda}$ pair production. 
		   }
		\end{figure}

	For $\Lambda \bar{\Lambda}$ and $p \bar{\Lambda}$ system,
	the two corresponding u-channel processes were simulated as the currently non-removable backgrounds. The event distributions of those two mechanisms dominate Sectors 6 and 2 on the right panel of Fig.~\ref{fig:vhplots}.
	The $\Lambda\bar{\Lambda}$ invariant mass and $t-t_{\texttt{min}}$ distributions of selected GlueX data ($\omega < 210^{\circ}$) are compared to the simulation (Fig.~\ref{fig:lamlambar_MC}). 
	The simulation combining the t-channel (signal) and u-channel (background) agrees with GlueX data.
	
	This result shows that our phenomenological model containing four different production mechanisms can be tuned to match the GlueX data distributions,
	with several parameters such as $\texttt{m}_{\texttt{BW}}, ~~\Gamma_{\texttt{BW}}, ~~b_R$ in Monte Carlo simulation. 
	In the future, the simulation together with multivariate analysis such as boosted decision tree technique (BDT) will be useful to further separate u-channel from t-channel events.

\section{Conclusion}

This article has shown preliminary result from GlueX data in photoproduction reactions $\gamma p \rightarrow p\bar{p} p $ and $\gamma p \rightarrow \Lambda  \bar{\Lambda}  p$. From the van Hove analysis of both reactions, it appears that multiple production mechanisms are present. To further understand the two reaction channels, a simulation of several possible production mechanisms were tuned to reasonably approximate the data distributions and some comparisons have been shown here.

A further goal of the study is to investigate the angular momentum structure of strangeness production through the study of spin correlations between the $\Lambda$ hyperons. Using linearly polarized photons peaking near $9$ GeV, observables such as the beam spin asymmetry can be studied, and will be presented at future conferences.

\section{Appendix}

	The radial ($\rho$) and angular ($\omega$) coordinates in van Hove diagram (Fig.~\ref{fig:vanhove}) can be expressed explicitly in terms of the three-body longitudinal momentum components as:
		\begin{align}
			\omega &= \arctan{\Bigg(\frac{-\sqrt{3} p_z^{B}}{p_z^{B}+2p_z^{\bar{B}}}\Bigg)}+\pi\\
			\rho &=\sqrt{\frac{2}{3}}\sqrt{[p_z^{B}]^2 + [p_z^{\bar{B}}]^2 + [p_z^{p}]^2}.
		\end{align}

\begin{acknowledgments}
We wish to acknowledge the support of Dr. Naomi Jarvis in setting up the analysis computing environment. 
We also wish to thank Carnegie Mellon undergraduate students Mr. Viren Bajaj and Mr. Samuel Dai for their efforts in the early stage of this project.
The work of the Medium Energy Physics group at Carnegie Mellon University was supported by
DOE Grant No. DE-FG02-87ER40315. The Thomas Jefferson National Accelerator Facility is supported by the U.S.
Department of Energy, Office of Science, Office of Nuclear Physics under contract DE-AC05-06OR23177.
\end{acknowledgments}

\nocite{*}
\bibliography{MENU19}

\end{document}